\documentclass[useAMS,usenatbib]{mn2e}
\usepackage{epsfig}
\usepackage{amssymb}
\usepackage{amsmath}

\def\arcsec{$^{\prime\prime}$}

\title[The Link Between SCUBA and \textit{Spitzer}: Cold
Galaxies at z$\lesssim$1]{The Link Between SCUBA and \textit{Spitzer}: Cold Galaxies at z$\lesssim$1}
\author[Symeonidis et al.]{Symeonidis,
M.$^{1}$\thanks{msy@mssl.ucl.ac.uk}, Page, M. $^{1}$,
Seymour N.$^{1}$, Dwelly T.$^{2}$, Coppin, K.$^{3}$, McHardy, I.$^{2}$, \and
Rieke, G. H.$^{4}$ and Huynh, M.$^{5}$ \\
$^{1}$University College London, Mullard Space Science Laboratory,
Holmbury St. Mary, Dorking, Surrey RH5 6NT, UK\\
$^{2}$ School of Physics $\&$ Astronomy, University of Southampton,
Southampton, Hampshire SO17 1BJ, UK \\
$^{3}$ Institute for Computational Cosmology, Durham University, South Road, Durham DH1 3LE, UK\\
$^{4}$ Department of Astronomy, The University of Arizona, Tucson AZ
85721, USA\\
$^{5}$ Spitzer Science Center, California Institute of Technology, 1200
East California Boulevard, Pasadena, CA 91125, USA }

\begin{document}

\date{Accepted  Received; in original form}

\pagerange{\pageref{firstpage}--\pageref{lastpage}} \pubyear{2009}

\maketitle

\label{firstpage}

\begin{abstract}

We show that the far-IR properties of distant Luminous and Ultraluminous InfraRed
 Galaxies (LIRGs and ULIRGs) are on average divergent from analogous sources in the local Universe.
Our analysis is based on \textit{Spitzer} MIPS and
 IRAC data of L$_{IR}$$>$10$^{10}$L$_{\odot}$, 70\,$\mu$m-selected objects in the 0.1$<$z$<$2 redshift range and supported by
 a comparison with
 the \textit{IRAS} Bright Galaxy Sample. The majority of the objects in our
 sample are described by Spectral Energy Distributions (SEDs) which peak at
 longer wavelengths than local sources of equivalent total infrared
 luminosity. This shift in SED peak wavelength implies a noticeable change in the dust and/or star-forming
 properties from z$\sim$0 to the early Universe, tending towards
 lower dust temperatures,
 indicative of strong evolution in the cold dust, `cirrus', component. We show that
 these objects are potentially the
 missing link between the well-studied local IR-luminous galaxies, \textit{Spitzer} IR
 populations and SCUBA sources --- the
 z\,$<$\,1 counterparts of the cold z\,$>$\,1 SubMillimetre Galaxies (SMGs) discovered in
 blank-field submillimetre surveys. The \textit{Herschel Space
 Observatory} is well placed to fully characterise the nature of these objects, as its coverage extends over a major part of the far-IR/submm
 SED for a wide redshift range.

\end{abstract}

\begin{keywords}
galaxies: general
galaxies: high-redshift
galaxies: starburst
galaxies: photometry
infrared: galaxies
\end{keywords}

\section{Introduction}
\label{sec:intro}

In the past 30 years, the field of infrared astronomy has instigated
immense breakthroughs in our understanding of the cosmos, at
the same time driving pioneering technological
advances (see review by Low, Rieke $\&$ Gehrz
2007\nocite{2007ARA&A..45...43L}). 
Discovering and resolving a large part of the Cosmic InfraRed Background
(CIRB) (Stanev $\&$ Franceschini 1998\nocite{1998ApJ...494L.159S}, Gorjian, Wright $\&$
Chary 1999\nocite{1999AAS...195.1402G}) led to valuable insights on the nature of the sources that are
responsible for more than half the cosmic electromagnetic energy
density originating from star-formation and Active Galactic Nuclei (AGN). Only 1/3
of the CIRB is attributed to the local galaxies discovered in abundance
by the \textit{InfraRed Astronomical Satellite} (\textit{IRAS}), which implies that a
major part of it is composed of redshifted infrared light originating from
galaxies at high redshift. Moreover, the CIRB intensity and peak
wavelength suggest that
the main contributors are dust-obscured populations whose bolometric
energy output is dominated by emission in the infrared. As studies have repeatedly
confirmed, these galaxies hide an immensely active InterStellar Medium (ISM); they are the
ultimate stellar nurseries, many also hosting deeply embedded and
rapidly growing AGN (e.g. Soifer et al. 1986,
Sanders et al. 1987, Genzel et al. 1998, Tacconi et al. 2002). 

Following these discoveries, a classification scheme was introduced based
on the total
infrared luminosity (L$_{IR}$) in the 8--1000\,$\mu$m spectral region and four main
classes of IR-luminous galaxies were defined: StarBursts
(SBs, 10$^{10}$\,$<$\,L$_{IR}$\,$<$\,10$^{11}$\,L$_{\odot}$), Luminous InfraRed
Galaxies (LIRGs, 10$^{11}$\,$<$\,L$_{IR}$\,$<$\,10$^{12}$\,L$_{\odot}$),
UltraLuminous InfraRed Galaxies (ULIRGs,
10$^{12}$\,$<$\,L$_{IR}$\,$<$\,10$^{13}$\,L$_{\odot}$) and HyperLuminous
InfraRed Galaxies (HyLIRGs, L$_{IR}$\,$>$\,10$^{13}$\,L$_{\odot}$), e.g. Sanders $\&$ Mirabel (1996)\nocite{1996ARA&A..34..749S}, Genzel et al. (1998)\nocite{1998ApJ...498..579G}. This classification is non-arbitrary and a good
representation of the generic characteristics of these sources, since the
transition in energetic output from 10$^{10}$ to 10$^{14}$L$_{\odot}$ is a
consequence of a change in dust properties (e.g. Noll et al. 2006), star formation history (e.g. Cole et
al. 2000\nocite{2000MNRAS.319..168C}, Brinchmann $\&$ Ellis
2000)\nocite{2000ApJ...536L..77B}, stellar mass (e.g. Bundy et al
2006\nocite{2006ApJ...651..120B}), morphology (e.g. Murphy et
al. 1996\nocite{1996AJ....111.1025M}, Wang et
al. 2006\nocite{2006ApJ...649..722W}) and the role of the central
black hole (e.g. Veilleux, Sanders $\&$ Kim
1997\nocite{1997ApJ...484...92V}, Brand et
al. 2006\nocite{2006ApJ...644..143B}).

With data from \textit{Spitzer}, doors to the high redshift Universe were
opened, both in terms of resolving the CIRB (e.g. Lagache, Puget $\&$
Dole 2005\nocite{2005ARA&A..43..727L}, Dole et al. 2006) and accurately pinpointing the evolution of the infrared luminosity
function (e.g. Le Floch' et al. 2005\nocite{2005ApJ...632..169L}, Huynh et
al. 2007\nocite{2007ApJ...667L...9H}), as well as identifying the key
processes responsible for the extreme luminosities of these galaxies
(e.g. Franceschini et al. 2003\nocite{2003A&A...403..501F}).
In terms of extragalactic science and specifically deep \textit{Spitzer} surveys, the capabilities of MIPS have been
extensively exploited in the mid-IR, enabling the nature of the 24\,$\mu$m
population to be characterised up to z$\sim$3 (e.g. Chary et al. 2004\nocite{2004ApJS..154...80C},
Marleau et al. 2004\nocite{2004ApJS..154...66M}, Le Floch et al. 2004\nocite{2004ApJS..154..170L}; 2005\nocite{2005ApJ...632..169L}, Houck et al. 2005\nocite{2005ApJ...622L.105H}, Dole
et al. 2006\nocite{2006A&A...451..417D}, Marcilac et al. 2006\nocite{2006A&A...451...57M}). Work on the 160\,$\mu$m population has been limited to
source count analysis (Dole et al. 2004a\nocite{2004ApJS..154...87D}, Frayer et al. 2006a\nocite{2006AJ....131..250F}), whereas
70\,$\mu$m-selected sources have also been characterised with respect to
their IR colours, mid-IR spectra and the
infrared luminosity function (Dole et
al. 2004a\nocite{2004ApJS..154...87D}, Frayer et
al. 2006a\nocite{2006AJ....131..250F};
2006b\nocite{2006ApJ...647L...9F}, Huynh et
al. 2007\nocite{2007ApJ...667L...9H}, Brand et
al. 2008\nocite{2008ApJ...673..119B}). 

The difference in resolution and sensitivity between the three MIPS bands has not
enabled a true unification of 24, 70 and 160\,$\mu$m populations and hence
the range and types of galaxies that dominate each one. For example, all \textit{Spitzer} sources detected at 70\,$\mu$m have 24\,$\mu$m counterparts, but the
reverse is not true, as the sensitivity of MIPS at 24\,$\mu$m is up
to a factor of 100 higher than at 70\,$\mu$m, considerably greater than the f$_{70}$/f$_{24}$ flux
density ratio for the average IR-luminous galaxy at z$>$1 (typically
$<$\,30, Papovich et al. 2007). In addition, there is a wide range of sources
with weak far-IR emission that
would form part of a 24\,$\mu$m survey, but would not necessarily have
detections in the MIPS 70 or 160\,$\mu$m
bands, such as quiescently star-forming
spirals, low redshift early-type galaxies, optically-selected quasars
and obscured AGN (e.g. Yan et al. 2004\nocite{2004ApJS..154...60Y}, Desai et al. 2008\nocite{2008ApJ...679.1204D}). 
On the contrary, due to the lower sensitivity of MIPS at 70\,$\mu$m, the
majority of far-IR selected objects conform to one main type and are thus defined by large amounts of dust, high IR to optical ratios and
high star formation rates; they are predominantly SBs, LIRGs and ULIRGs
(Symeonidis et al. 2007; 2008). This brings a high degree of homogeneity in
the 70\,$\mu$m population, clearly benefiting any study of its evolution with redshift. 

Infrared Spectral Energy Distributions (SEDs) for \textit{Spitzer}
selected sources have been examined with and without individual 70 and
160\,$\mu$m detections, the latter via 70 and 160\,$\mu$m stacking on the 24\,$\mu$m
positions (Papovich et al. 2007\nocite{2007ApJ...668...45P}, Zheng
et al. 2007\nocite{2007ApJ...670..301Z}, Bavouzet et al. 2008\nocite{2008A&A...479...83B}) and the former with data from various
surveys, such as the \textit{Spitzer} Infrared Nearby Galaxies
Survey (SINGS) (Kennicutt et al. 2003\nocite{2003PASP..115..928K}), the \textit{Spitzer} Wide-area InfraRed
Extragalactic survey (SWIRE) (Lonsdale 2003\nocite{2003PASP..115..897L}; 2004\nocite{2004ApJS..154...54L}), the All-wavelength Extended Groth Strip
Survey (AEGIS) (Davis et al. 2007\nocite{2007ApJ...660L...1D}), the Bootes field of the NOAO
Deep Wide Field Survey (NDWFS) and the \textit{Spitzer} First Look Survey
(FLS) (see Dale et al. 2005\nocite{2005ApJ...633..857D}, Rowan-Robinson et al. 2005\nocite{2005AJ....129.1183R}, Symeonidis et
al. 2007\nocite{2007ApJ...660L..73S}; 2008\nocite{2008MNRAS.385.1015S},
Bavouzet et al. 2008\nocite{2008A&A...479...83B} for work on SEDs).
However, due to \textit{Spitzer}'s limited sensitivity and angular resolution at
the long wavelengths, the exact shape of the SED for various galaxy types has not been
examined to the detail merited for the high redshift Universe. For non-local sources, the short
wavelength (Wien) side of the SED peak has been mapped in the mid-IR and far-IR
up to z$\sim$2 and z$\sim$1, respectively, whereas submm studies have
examined the long wavelength (Rayleigh-Jeans) side up to
z$\sim$3.  Although it seems that cumulative data over such a wide
wavelength and redshift range would have enabled the complete
characterisation of the IR SED, the different sensitivities of mid-IR, far-IR and submm surveys
do not favour homogeneneous  investigations and as a result, many details
on the SED are missing. Moreover, these `bolometric' studies
do not provide detailed information on the key
intragalactic environments which dominate the energy budget and as a result, fundamental
questions with respect to the evolution of the properties of obscured
galaxies still remain unanswered.

The natural question that follows is whether local
infrared galaxies are the same, similar for the most part or instead very different to their high
redshift counterparts. Here, we approach this topic by
comparing the SEDs of local (z$\sim$0) infrared galaxies with galaxies
at z$>$0.1, selected at 70\,$\mu$m, based on the preliminary results of Symeonidis
et al. (2008\nocite{2008MNRAS.385.1015S}, hereafter S08). In section 2 we define the
sample and outline the various selection criteria. Sections 3 and 4 form
the main body of our analysis, where we describe our methodology and
results, paying particular attention to any biases that could have been
introduced. In section 5 we discuss the implication of our results,
making comparisons to other studies. Finally, the summary and conclusions are presented in section 6. Throughout we use a concordance model of Universe expansion, $H_0=70$ km\,s$^{-1}$Mpc$^{-1}$, $\Omega_M=0.3$ and
$\Omega_{\Lambda}=0.7$ (Spergel et al. 2003).

\section{Sample Selection}
\label{sec:sample}

This work is based on Guaranteed Time Observations (GTO) using the
Multiband Imaging Photometer (MIPS, Rieke et
al. 2004\nocite{2004ApJS..154...25R}) on the \textit{Spitzer Space Telescope}
(Werner et al. 2004\nocite{2004ApJS..154....1W}) in
the EGS field ($\sim$0.5\,deg$^2$) (Davis et al. 2007) and 13$^H$ XMM-Newton/Chandra Deep Field ($\sim$0.6\,deg$^2$, centred on
$13^h\,34^m\,37^s$\,+\,$37^{\circ}54'44''$, Seymour et al. 2008\nocite{2008MNRAS.386.1695S}).
We select the sample at 70$\mu$m and retrieve 178 objects down to
$\sim$\,4\,mJy (5$\sigma$) in the EGS and 244 objects down to
$\sim$8\,mJy (5$\sigma$) in 13$^H$ Field. As these two surveys cover
similar areas and share a similar photometric completeness limit at f$_{70}$\,$\sim$\,10\,mJy, we are able to combine the data into
one congruous sample, where 4\,$<$\,f$_{70}$\,$<$\,290\,mJy. 

\subsection{Source extraction and photometric coverage}

For details on MIPS data reduction and source extraction, we refer the reader to Symeonidis et al. (2007\nocite{2007ApJ...660L..73S})
and Seymour et al. (2009), for the EGS and the 13$^H$ field
respectively. Each 70\,$\mu$m object was matched to the brightest
24\,$\mu$m source within 8\arcsec, in accordance with the MIPS PSF. Given the relative sensitivity limits
($\sim$\,110$\mu$Jy for 24$\mu$m, $\sim$\,10\,mJy for 70$\mu$m), two
sources near the 70$\mu$m 5$\sigma$ flux density
limit with no 24\,$\mu$m counterparts, were considered spurious and discarded. 
The 160\,$\mu$m photometry was extracted with the IRAF/DIGIPHOT package on the known positions of
the 70\,$\mu$m sources, enabling us to go nearer the confusion limit,
estimated to be on the order of $\sim$\,40\,mJy at 160\,$\mu$m (e.g. Dole,
Lagache $\&$ Puget 2003\nocite{2003ApJ...585..617D}). We used an aperture radius of 40\,\arcsec
(equivalent to the 160\,$\mu$m diffraction limit) and the
aperture corrections specified by the MIPS team (1.884 for a 40\arcsec
radius and 40-75\arcsec background annulus;
http://ssc.spitzer.caltech.edu/mips/apercorr/). We tested our method of source extraction on the
Spitzer Cosmic Evolution Survey (S-COSMOS, Frayer et
al. 2009\nocite{2009arXiv0902.3273F}) 160$\mu$m
image, with the known positions of the S-COSMOS 160$\mu$m sources and
found our flux density estimates to be consistent, within
10 per cent (r.m.s.) of the S-COSMOS MOPEX (Makovoz $\&$ Marleau
2005\nocite{2005PASP..117.1113M}) estimates.

The MIPS photometry was supplemented with InfraRed Array Camera (IRAC) 8\,$\mu$m data to ensure a more
extensive coverage of the infrared SED. The overlapping area of the
8, 24 and 160\,$\mu$m images adds up to about 65 per cent of the EGS
70\,$\mu$m survey and 54 per cent of the 13$^H$
field 70\,$\mu$m survey. Requiring IRAC coverage results in a subset of
114 70\,$\mu$m sources in the EGS and 132 in the 13$^H$ field, instead
of the 178+224 described earlier.

\subsection{Redshifts for the 13$^H$ field}
\label{13hr_redshifts}

Photometric redshifts were estimated for the 13$^H$ field, using multi-band deep
optical and near-IR imaging: $u^*$, $B$, $g'$, $R$, $i'$, $I$, $z'$, $Z$,
$J$, $H$, and $K$, from various ground based observatories (CFHT, Subaru,
INT and UKIRT) and 3.6, 4.5, 5.8, 8.0$\mu$m from \textit{Spitzer}'s IRAC. The
requirement of area coverage in at least 4 optical/near-IR bands, reduced the
number of available 70\,$\mu$m sources from 132 to
108. The cross-matching was
done using the 24\,$\mu$m positions each 70\,$\mu$m
source was identified with. Photometric redshifts were estimated with
the HyperZ (Bolzonella et al. 2000\nocite{2000A&A...363..476B}) template fitting code, including
the set of galaxy and AGN SED templates from Rowan-Robinson et al.
(2008) and setting optical extinction as a free variable. All
aperture magnitudes were also corrected for aperture losses. Moderate optical
extinction was permitted for the AGN and late type galaxy templates, and
heavy extinction (up to $A_V = 10$) was allowed for the starburst
templates. For a full description of the photomety extraction and
redshift fitting process see Dwelly et al. (in prep.). 

Within the final overlap area of the \textit{Spitzer} MIPS/IRAC and
ground-based optical/near-IR surveys, 39 good quality spectroscopic redshifts are also available, as part of an ongoing program to identify the optical
counterparts of faint X-ray and radio sources in the 13$^H$ field, using various long-slit and multi-object spectrographs
at the WHT, CFHT and Keck telescopes. The reliability
of the photometric redshifts was checked against the spectroscopic
redshifts, finding high consistency with an r.m.s. scatter in redshift
of only $\sim$\,0.05.

Finally, as the aim of this work is to evaluate the properties of non-local
70\,$\mu$m selected sources, we made a redshift cut at z=0.1, rejecting
the 26 low redshift resolved bright galaxies in the 13$^H$ field.
The final 13$^H$ field redshift sample consists of 82 sources, 34
with spectroscopic redshifts and 48 with photometric redshifts.

\begin{figure}
\epsfig{file=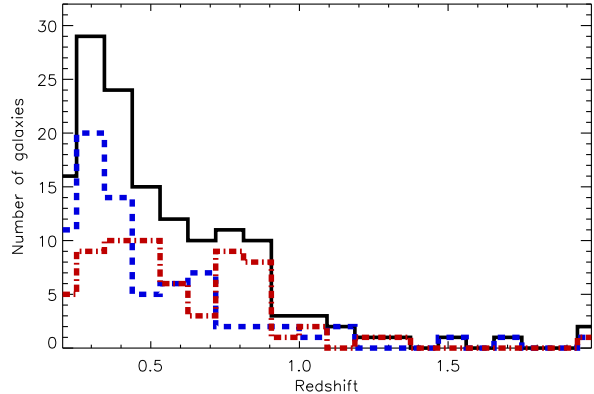,width=8.7cm}
\caption{The combined redshift distribution of the \textit{70$\mu$m
 sample} (z$_{mean}$\,$\sim$\,0.5, z$_{median}$\,$\sim$\,0.42; solid black line) from the individual distributions of the EGS
 population (dot-dashed red line) and the 13$^H$ field population (dashed
 blue line). The difference between the two redshift distributions is likely due to sample variance.}
\label{fig:distz}
\end{figure}

\subsection{Redshifts for the EGS}

Spectroscopic redshifts for the EGS were acquired by the Deep Extragalactic Evolutionary Probe 2
(DEEP\,2) survey (Davis et al. 2003\nocite{2003SPIE.4834..161D}). DEEP\,2 targeted sources in the
EGS with a magnitude limit of R$_{AB}$\,=\,24.1. The brightest, most extended
sources were weeded out of the spectroscopic targeting algorithms,
as their photometry was suspect and they could potentially saturate the
detector. Thereafter, the retrieval of redshifts was random, at $\sim$60$\%$ sampling and at the time of
writing there are 66 good quality redshifts available for the 114
sources in the MIPS/IRAC overlap areas. These are in the range 0.1\,$<$\,z\,$<$\,1.9,
with a mean of 0.55 and a median of 0.48. The exclusion of the brightest R$<$18 
sources implies that there is an under-representation of extended,
low-redshift galaxies. However, as we apply a
redshift cut at z=0.1 (see section \ref{13hr_redshifts}), these would
have been eliminated from our final sample anyway.
On the faint end, there are only 4
objects fainter than R=24.1 in the full sample, i.e. only two
would have made it in the redshift sample, so overall there is
high consistency between the two samples.

The limited optical
photometry for the EGS 70\,$\mu$m objects did not allow us to calculate reliable
photometric redshifts and as a result, we only 
use these 66 sources with spectroscopic redshifts in subsequent
work. However, as outlined above, this redshift
sample is a good represenation of the the z$>$0.1 parent EGS
sample.

\subsection{The final 70\,$\mu$m sample}

All 148 sources, assembled after the various area and redshift cuts
described in earlier sections, are detected at 8 and
24\,$\mu$m. Eight sources have signal to noise $<$\,1 at 160\,$\mu$m and are therefore
excluded from subsequent analysis.
From the remaining 160\,$\mu$m measurements, $\sim$\,80 per cent
are $>$3$\sigma$ detections, $\sim$\,62 per cent $>$5$\sigma$ and $\sim$\,47 per cent
$>$7$\sigma$. 

The final sample consists of 140 galaxies in the 0.1--2 redshift
range (z$_{mean}$\,$\sim$\,0.5, z$_{median}$\,$\sim$\,0.42, hereafter
the \textit{70\,$\mu$m sample}). 
Figure \ref{fig:distz} shows the individual and combined redshift
distributions, the differences likely arising due to sample variance.

Completeness is not a strict requirement for our analysis. However the selection criteria we apply, ensure
that the \textit{70\,$\mu$m sample} is representative of the parent 70\,$\mu$m population. 
Nonetheless, given the strong correlation between R-band magnitude and redshift, we
examine the possibility that the requirement of (optically
derived) redshifts modifies the type of 70\,$\mu$m sources present in the final
sample. We find that the redshift at which each source becomes too
faint for inclusion in the 70\,$\mu$m sample is much lower than the
redshift at which it drops out of the R-band survey. 
This is to be expected as the R-band limit ($<$\,1\,$\mu$Jy), is 4-5 orders of magnitude lower than the 70\,$\mu$m
flux density limit ($\sim$\,10\,mJy), but at the same time $\sim$\,95
per cent of the objects have f$_{70}$/f$_R$\,$<$5000. Under these circumstances, sources
would drop out of the 70\,$\mu$m selection before dropping out of the
optical selection, confirming that the sample of 140
sources we focus on, is representative of z$>$0.1 70\,$\mu$m detected galaxies.

\subsection{The local sample}
In order to compare local and high redshift sources in a way that is
as unbiased as possible, we evaluate our results against 259 galaxies from the
\textit{IRAS} revised Bright Galaxy Sample (BGS) selected at 60\,$\mu$m (hereafter,
the \textit{local sample}) by using available photometry from Sanders et
al. (2003, hereafter S03)\nocite{2003AJ....126.1607S}. The \textit{IRAS} BGS is
particularly suitable for such a comparison because it includes all nearby LIRGS and
ULIRGs and the 60\,$\mu$m selection criterion is comparable to
70\,$\mu$m at z$\sim$0.1 where our sample begins. We do not weed out any
sources which potentially have a significant AGN contribution from either the \textit{local} or the
\textit{70\,$\mu$m sample}. Apart from the fact that there is a low AGN incidence in
far-IR selected samples, both populations
are defined by powerful far-IR
emission, hence the presence of large amounts of dust, explicitly
setting star-formation to be a very a significant
part of the galaxies' energetics, if not the dominant. The \textit{IRAS} BGS consists of
629 galaxies (z\,$\lesssim$\,0.08), the majority of which are starbursts, but with a significant
LIRG population of $\sim$\,25 per cent and a very small fraction of ULIRGs
($\sim$\,2 per cent). We only use galaxies
which are not flagged as having large uncertainties in the photometry:
all the available LIRGs and ULIRGs (143 and 18) and 98
randomly-selected starbursts. As our main conclusions
rest on the comparison between local and high redshift LIRGs and ULIRGs and
much less so on SBs, we believe it unnecessary to include all of the \textit{IRAS} SBs in the
\textit{local sample}. Consequently, we rely on a randomly-selected subset, which is about
3-4 times the number of starbursts in the \textit{70\,$\mu$m sample}, to represent the BGS SBs.

\section{SED fitting}
\label{sec:fitting}

The SED shape of an IR-luminous galaxy in the 5--1000\,$\mu$m range
(rest wavelength), is dominated by
emission from dust, with a range of temperatures spanning $\sim$\,3 orders of
magnitude from $\sim$\,1500\,K at the shortest wavelengths to $\sim$\,10\,K
at the longest. In the near-IR ($\sim$\,2--5\,$\mu$m), light from
stellar photospheres dominates the continuum, although there can be a small contribution from dust (e.g. Lu et
al. 2003\nocite{2003ApJ...588..199L}), subsequently becoming negligible
shortward of 2\,$\mu$m, as the maximum grain sublimation temperature is of the order of $\sim$\,1500\,K.
The mid to far infrared part of the SED has two major contributions: emission
from stochastically heated Very Small Grains (VSGs), with T$>$\,60\,K, responsible for the continuum up to about
$\lambda$\,$\sim$\,40\,$\mu$m and emission from large grains in thermal
equilibrium, with T$<$\,50\,K, 
responsible for emission longward of $\lambda$\,$\sim$40\,$\mu$m (e.g. Efstathiou $\&$ Rowan-Robinson 1990\nocite{1990MNRAS.245..275E};
1995\nocite{1995MNRAS.273..649E}, Granato $\&$ Danese
1994\nocite{1994MNRAS.268..235G}, Silva et
al. 1998\nocite{1998ApJ...509..103S}, Klaas et
al. 2001\nocite{2001A&A...379..823K}). Accordingly, the rest-frame
SED peak, which defines the wavelength of maximum
energetic output (in $\nu$f$_{\nu}$), is in the 40--140\,$\mu$m range and the SED is often approximated by a
modified black body ($B_{\nu} \nu^{\beta}$) of T$\sim$\,20--60\,K and emissivity $\beta$ $\sim$
1--2. As the bulk of the dust mass is in equilibrium, such an
approximation does represent the average SED shape, however a colder, `cirrus' component associated with the
properties and distribution of dust in the ISM and the intensity of the
diffuse radiation field is often missed. If the cold component is significant,
yet ignored, the integrated energy
output of the galaxy will be underestimated, but more importantly it
will lead to erroneous estimates for physical
quantities, such as dust mass, emissivity, opacity and
temperature ---  e.g. see Silva et al. (1998\nocite{1998ApJ...509..103S}), Granato et
al. (2000\nocite{2000ApJ...542..710G}) and Siebenmorgen, Krugel $\&$ Laurejis
(2001\nocite{2001A&A...377..735S}) where the importance of both the
starburst and cirrus components is emphasized.  

In this paper, we broadly categorise the sources as `cold' or `warm', depending on
the average temperature that defines the bulk of the dust mass, which,
as mentioned earlier, is
mostly found to be in thermal equilibrium. As a result the average dust
temperature scales with the inverse of the SED peak wavelength; a cold galaxy SED is
expected to peak at longer wavelengths, whereas a warm SED will peak at shorter wavelengths.
Note that this is not the same as the classical \textit{IRAS} `warm/cold' terminology
parametrised as the mid-IR $f_{25}/f_{60}$ continuum slope. The \textit{IRAS}
`warm/cold' $f_{25}/f_{60}$ criterion is based on the expectation that dust heated by an AGN
would reach higher temperatures and hence lower the continuum slope between 25 and
60\,$\mu$m (or 24 and 70\,$\mu$m in \emph{Spitzer}'s case) (e.g. Farrah et
al. 2005\nocite{2005ApJ...626...70F}; Verma et al. 2005\nocite{2005SSRv..119..355V}, Frayer et
al. 2006\nocite{2006AJ....131..250F}); it has been applied
extensively in order to quantify relative AGN/starburst contributions in
ULIRGs (e.g. de Grijp et al. 1985\nocite{1985Natur.314..240D}, Miley, Neugebauer $\&$ Soifer
1985\nocite{1985ApJ...293L..11M}, Soifer et
al. 1989\nocite{1989AJ.....98..766S}).
 What we refer to as cold/warm, is not related to the
mid-IR continuum slope, but solely to the far-IR properties and
more specifically, the position of the SED peak.

\begin{figure}
\epsfig{file=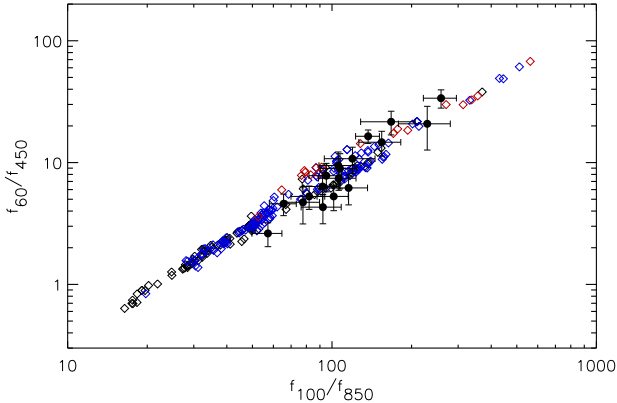,width=8.7cm}
\caption{The 60/450 vs 100/850 flux density ratio for local galaxies. Open diamonds: predicted colours for the
 \textit{local sample} using the SK07 fits
 (black for SBs, blue for LIRGs and red for ULIRGs). Filled black
 circles: observations from Dunne $\&$ Eales (2001, DE01). Note the good agreement between predicted and observed colours, verifying both
 the accuracy of our fitting method and the validity of the SK07 models.}
\label{fig:newsubmm450}
\end{figure}

\subsection{Methodology}
\label{sec:methodology}

We fit the available IR data for the \textit{70\,$\mu$m} and \textit{local sample} with the Siebenmorgen $\&$ Krugel
(2007\nocite{2007A&A...461..445S}, hereafter SK07) \emph{theoretical} SED
templates. In S08, we found that these models are optimum for fitting a
wide range of source types, as their
formulation employs a configuration of stars and dust distributed
throughout the volume. In the SK07 formulation, mid-IR emission originating in dust regions enveloping OB
stars is deconvolved from the cirrus emission due to the general stellar radiation
field, allowing the far-IR part of the SED to be modeled as a two-temperature
component; see also Klaas et al. (2001\nocite{2001A&A...379..823K}),
Dunne and Eales (2001\nocite{2001MNRAS.327..697D}) for the benefits of a
2-temperature far-IR formulation. 
The SK07 library consists of $\sim$\,7000 templates defined by the physically acceptable combinations
of the following 5 parameters: radius of dust emitting region
(0.35, 1 and 3 kpc), visual extinction (A$_V$\,=\,2–-150), total infrared
luminosity (10$^{10}$--10$^{14.7}$\,L$_{\odot}$), hot spot dust density (100-–10000
cm$^{-3}$) and  L$_{stars, OB}$/L$_{stars, tot}$ (40, 60 and 90 per cent). Hot spot dust
density refers to the density of the dust region enveloping OB
stars. L$_{stars, OB}$/L$_{stars, tot}$, represents the percentage of the total
infrared luminosity that originates from OB stars, compared to the contribution from the
general stellar population. 

We perform a $\chi^2$ fit on the photometry (8--160\,$\mu$m for the
\textit{70\,$\mu$m sample}; 12--100\,$\mu$m for the \textit{local
sample}), using the entire range of SK07 templates, folding in the photometric redshift
uncertainties (where applicable). For each object, we
determine one interesting parameter, namely the wavelength of the SED ($\nu$L$_{\nu}$)
peak, calculated using the mean (and 1$\sigma$ uncertainty) of all templates with $\chi_i^2
\le \chi^2_{min}+1$, where i is the template index and $\chi^2_{min}$ is
the minimum $\chi^2$ value. In the same way we also get estimates for the total infrared luminosity (L$_{IR}$, 8--1000\,$\mu$m)
and submm flux density at 350 and 850\,$\mu$m. 

We find that the \textit{70\,$\mu$m sample} consists of 30 SBs, 79 LIRGs, 29
ULIRGs and 3 HyLIRGs, indicating that down to the $\sim$\,10\,mJy level, L$>$10$^{10}$\,L$_{\odot}$
sources are the sole contributors to z$>$0.1 70\,$\mu$m
populations.

Note that any comparison between IR galaxies at various redshifts will
only be fair if each luminosity class is examined separately. As
outlined in section \ref{sec:intro}, the transition in luminosity
closely follows changes in the galaxies' physical properties,
star-formation history, etc., which must be reflected in all
comparative evaluations, something that we maintain throughout
this work.

\begin{figure}
\epsfig{file=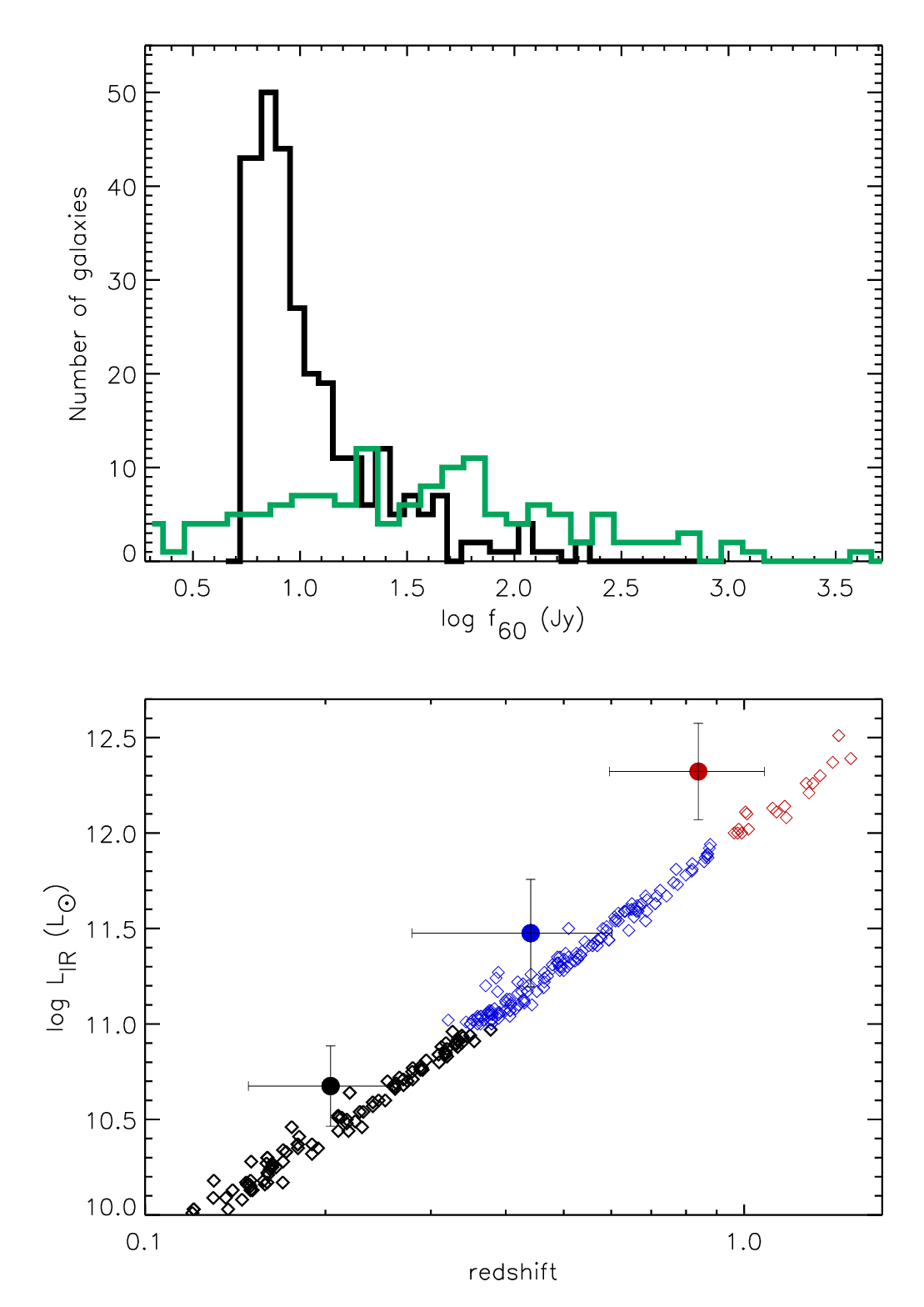,width=8.7cm}
\caption{{\bf{Top panel}}: The distribution in 60\,$\mu$m flux density for the \textit{local
 sample} (black histogram), compared with the distribution that the
 \textit{70\,$\mu$m sample} would have if it were at the average
 redshift of the local sample (z$\sim$0.012; $\sim$\,50\,Mpc) (green histogram). Note that apart from a few low
 luminosity SBs (L$<$5$\times$10$^{10}$\,L$_{\odot}$), the 70\,$\mu$m
 sample would be easily detectable within the 5.24\,Jy \textit{IRAS}
 flux density limit. {\bf{Lower panel}}: Total infrared luminosity
 plotted as a function of the maximum redshift at which
 the types of sources that form part of the \textit{local sample} would be detectable within the 70\,$\mu$m 10\,mJy flux density
 limit - SBs in black, LIRGs in blue and ULIRGs in red. Comparing with the
 average L$_{IR}$ and redshift for each luminosity class of the
 \textit{70$\mu$m sample} (filled circles), reveals that the 70\,$\mu$m
 selection is quite sensitive to the LIRGs and ULIRGs of
 the \textit{local sample}; at the high luminosity end, the sources in the \textit{local sample} can be detected to
higher redshifts than the average redshift of the 70\,$\mu$m galaxies. }
\label{fig:bias}
\end{figure}

\begin{figure*}
\epsfig{file=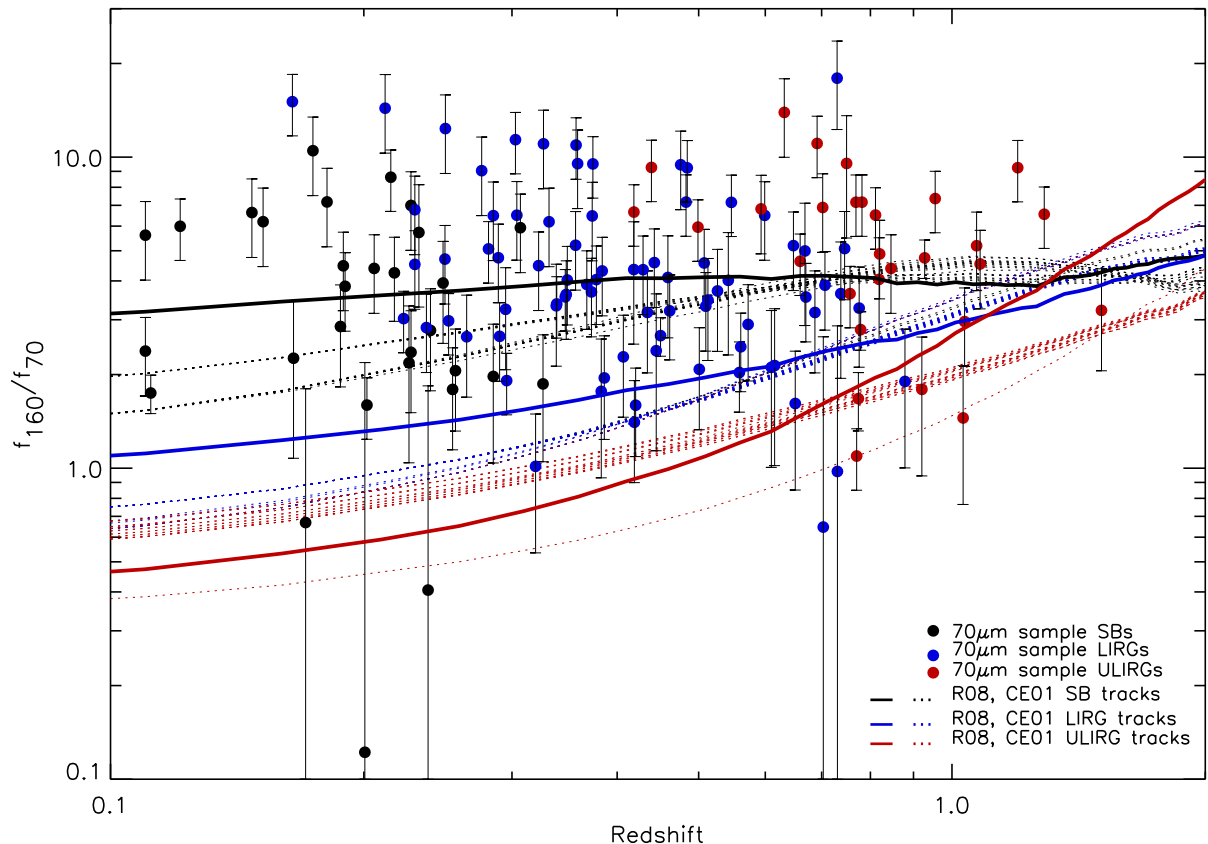,width=17cm}
\caption{The 160/70\,$\mu$m flux density ratio versus
 redshift for the SBs, LIRGs and ULIRGs of the \textit{70$\mu$m sample} (filled cirles). The dotted
 line tracks are the Chary $\&$ Elbaz (2001) templates and the solid line tracks
 are the Rieke et al. (2008) templates with luminosities of log L$_{IR}$
 = 10, 11, 12. All templates have been convolved with the MIPS 70 and 160\,$\mu$m filter
response curves, since MIPS photometry is calculated from the integrated flux in each band and is therefore not strictly
monochromatic. The colour coding for both data
 points and tracks is the same: black for starbursts, blue for LIRGs, red for ULIRGs. The errors have been
 estimated by combining calibration, photometric and statistical uncertainties. Note the large scatter in f$_{160}$/f$_{70}$ spanning
almost an order of magnitude within each luminosity class. Moreover, the
 f$_{160}$/f$_{70}$ values for many 70\,$\mu$m-detected LIRGs and ULIRGs are offset by at least a factor of 2 and up to 10 in extreme cases from local
 tracks of the same luminosity, suggesting that these sources are on average colder than their local counterparts.}
\label{fig:ratio70160}
\end{figure*}

\begin{figure*}
\epsfig{file=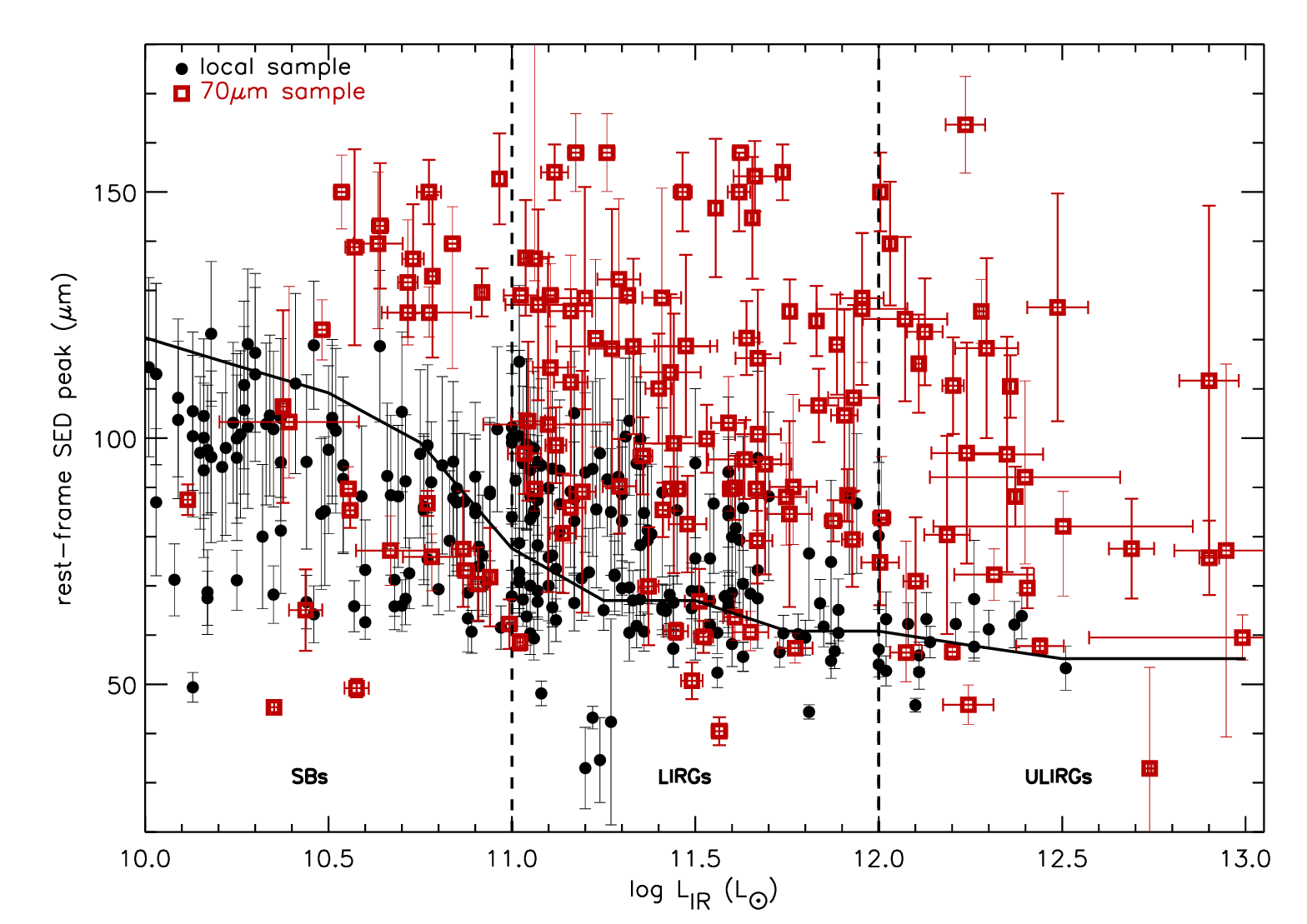,width=17cm}
\caption{The relationship between SED peak and total infrared
 luminosity for the \textit{local sample} (black filled circles) and
 \textit{70$\mu$m sample} (red open squares), split
 into 3 luminosity bins (SBs, LIRGs and ULIRGs), outlined by the
 vertical dashed lines. The solid line is the peak-L$_{IR}$ relationship of the Rieke
 et al. (2008) templates. Note that for a given L$_{IR}$, many high redshift
galaxies have on average significantly colder SEDs that peak at longer wavelengths.}
\label{fig:LIRpeak}
\end{figure*}

\subsection{Selection Issues}
\label{sec:biases}

\subsubsection{The SED models and fitting accuracy}
\label{sec:modelling}

We chose the SK07 models, as their theoretical formulation allows the
representation of a wide range of systems, minimising assumptions on the
types of sources that exist. This is in contrast to \emph{empirical} sets
of templates, e.g. Chary $\&$ Elbaz (2001), Dale $\&$ Helou
(2002\nocite{2002ApJ...576..159D}), which are tailored to local galaxies and could potentially
predispose the results towards the types of SEDs found locally. In S08
we show that this is indeed the case and find them un-representative of
the SEDs of 70\,$\mu$m-detected galaxies, especially in the far-IR. In
contrast, we find that the SK07 models are representative of the SEDs of
both the \textit{local} and \textit{70\,$\mu$m samples}. 

To quantify this further, we compare to submm data from Dunne et
al. (2000\nocite{2000MNRAS.315..115D}, hereafter D00) and Dunne $\&$
Eales (2001\nocite{2001MNRAS.327..697D}, hereafter DE01), who describe the submm properties of local
galaxies in the SCUBA Local Universe Galaxy Survey (SLUGS) and report
850 and 450\,$\mu$m observations for a fraction of the \textit{IRAS}
BGS. Using the SK07 models we predict 60, 100, 450 and 850\,$\mu$m flux
densities for the sources in the \textit{local sample} and evaluate the 
results against the observed DE01 far-IR/submm colours
(60/450 vs 100/850, figure \ref{fig:newsubmm450}). We find good agreement
between our predictions and the DE01 observations, which confirms both the
accuracy of our fitting method and the validity of the SK07 templates.

\subsubsection{The detection limits of the 60\,$\mu$m \textit{IRAS}
   survey and 70\,$\mu$m \textit{Spitzer} survey }
\label{sec:limits}

Selecting the sample at a wavelength shortward of the peak
(the Wien side of the SED), by definition, does not predispose the selection towards cold
sources, in the same way that selecting longward of the peak (the
Rayleigh-Jeans side) does not predispose towards warm sources. 
Our 70\,$\mu$m selection is well into the Wien side of the SED (for
z\,$>$\,0.2 it corresponds to $\lambda_{rest}$\,$<$\,58$\mu$m),
so any bias would favour the selection of warmer SEDs, especially for galaxies at
z\,$>$\,0.5. The 60\,$\mu$m \textit{IRAS} selection probes longer
wavelengths in local galaxies than the 70\,$\mu$m MIPS photometry for
sources at z$>$0.2, so, if anything, it should be
more sensitive to relatively cold SEDs. Also, note that the 160\,$\mu$m photometry coincides with the
\textit{IRAS} 100\,$\mu$m band for sources at 0.4$<$\,z\,$<$1; this applies to
all ULIRGs and the higher luminosity LIRGs in the \textit{70\,$\mu$m
sample}. 

We estimate the 60\,$\mu$m
flux density the 70\,$\mu$m sources would have if at the average redshift of the \textit{local sample} (z\,$\sim$\,0.012, $\sim$50\,Mpc) and
compare with the f$_{60}$ distribution for the \textit{local sample},
finding considerable overlap (figure \ref{fig:bias}, top panel). It is only the lowest luminosity 70\,$\mu$m-detected SBs
(L$<$\,5$\times$10$^{10}$L$_{\odot}$) that would not be recovered within
the \textit{IRAS} 60\,$\mu$m flux density limit of 5.24\,Jy. This is not surpising, as at 70\,$\mu$m we
detect them at the mJy level and at low redshifts (z$<$0.2). 
Using a (conservative) 70\,$\mu$m 10\,mJy flux density limit and the SK07 templates matched to the \textit{IRAS} galaxies, we calculate the maximum
redshift at which each source in the \textit{local sample} would still
be bright enough to appear in the 70\,$\mu$m survey (figure \ref{fig:bias}, lower panel). 
As expected, the types of sources that make up the \textit{local sample}
are easily recoverable at f$_{70}$\,$>$\,10\,mJy. In fact, at the high
luminosity end (log\,L$_{IR}$$\gtrsim$11.5), the sources in the \textit{local sample} are detectable to
higher redshifts than the average redshift of the galaxies in the \textit{70\,$\mu$m
sample}. 
The fact that the entire range of
template matches for L$>$\,5$\times$10$^{10}$L$_{\odot}$ sources in both samples is detectable with the
flux density limits of the Spitzer/MIPS and \textit{IRAS}
surveys, confirms that the samples are well-matched and that our results and conclusions (sections \ref{sec:results} and
\ref{sec:discussion}) are not affected by the initial selection. 

The final point to address is the dependency of the IR luminosity
and position of SED peak on the absolute value of the 160\,$\mu$m flux density. 
Our detection levels are at least 5\,$\sigma$ at 70\,$\mu$m and typically
4--5\,$\sigma$ at 160\,$\mu$m, so to simulate combined uncertainties we
evaluate the effects of a 30 per cent change in f$_{160}$ for all
objects in the \textit{70$\mu$m sample}. We find that the change in
luminosity does not exceed 0.2 dex and the average peak wavelength shift is less than 10\,$\mu$m, well within
the typical uncertainty (see section \ref{sec:peak}). This
confirms that the absolute value of f$_{160}$ does not drive our
estimates for $\lambda_{peak}$ and L$_{IR}$.

\begin{figure}
\epsfig{file=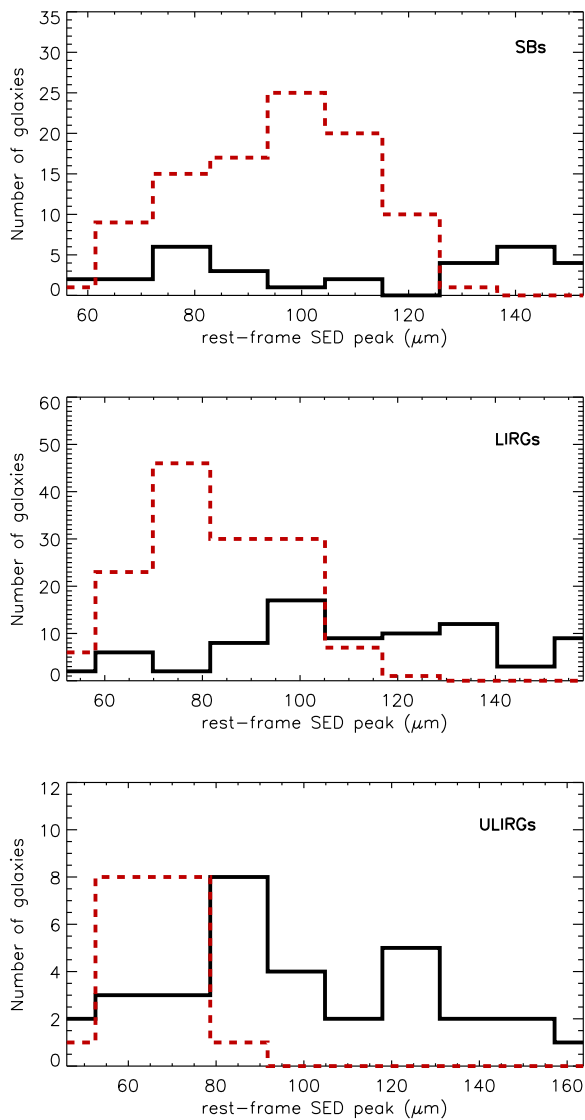,width=8.7cm}
\caption{The distribution in SED peak for the 3 main luminosity classes
 (SBs, LIRGs and ULIRGs). The dotted red line represents the
 \textit{local sample} and the solid black line represents the
 \textit{70$\mu$m sample}. Although there is overlap in the
 distributions, a significant fraction of high redshift LIRGs and
 ULIRGs peak at longer wavelengths than the local galaxies
 of the same luminosity.}
\label{fig:peakdist}
\end{figure}

\section{Results}
\label{sec:results}

\subsection{The 160/70$\mu$m flux density ratio}
\label{sec:70160}

As a first approach we examine the 
observed f$_{160}$/f$_{70}$ colours of the \textit{70\,$\mu$m sample} as a function
of redshift (figure \ref{fig:ratio70160}), relying on SED templates to represent the IR colours of local
galaxies. Accordingly, we chose the Chary $\&$ Elbaz (2001\nocite{2001ApJ...556..562C}, hereafter CE01) and Rieke et
al. (2008, hereafter R08) \emph{empirical} templates, well-representative of the average properties of
local galaxies, but with some key differences in the continuum slope
shortward of the peak; R08 consider their templates more accurate for
high luminosity IR galaxies, such as Arp220, as they
account for the steeper continuum slope seen in the spectra from the
InfraRed Spectrograph (IRS) on \textit{Spitzer}. 

The f$_{160}$/f$_{70}$ ratio directly scales with the
position of the peak, a relationship which is much clearer once
\textit{K}-corrections have been applied (see S08). Here, in order to minimise
assumptions, we only compare the observed f$_{160}$/f$_{70}$
colour against the locally-derived tracks. However, by splitting the
\textit{70\,$\mu$m sample} into luminosity classes we can evaluate
f$_{160}$/f$_{70}$ more consistently and in a relatively narrow redshift
range. 
The scatter in SB f$_{160}$/f$_{70}$ is in broad agreement with the
local SB SEDs, with a similar number of sources lying above and below
the region defined by the tracks. On the contrary, the f$_{160}$/f$_{70}$ for many LIRGs and
ULIRGs is displaced from the local LIRG and ULIRG tracks by at least a factor of 2 and
up to 10 in extreme cases. This suggests that a large fraction
of these galaxies are colder than their local equivalents, as the energetic
output at long wavelengths is increased relative to the short
wavelengths. 
Overall, the f$_{160}$/f$_{70}$ colour does not seem to change with
redshift and, although this would be mis-interpreted as a no-evolution
effect, it is in fact quite the opposite. Examining the
variation of f$_{160}$/f$_{70}$ with redshift will invariably have a
luminosity and \textit{K}-correction bias folded into it, as at each redshift range the survey
probes galaxies at different rest-frame wavelength and of different L$_{IR}$. In fact strong evolution is evident, as the high redshift LIRGs and
ULIRGs have f$_{160}$/f$_{70}$ ratios comparable to those of local low
luminosity starbursts as opposed to local LIRGs and ULIRGs.

\subsection{The SED peak}
\label{sec:peak}

In figure \ref{fig:LIRpeak}, the variation of SED peak (rest-frame wavelength) is plotted
against total
infrared luminosity for the \textit{local} and
\textit{70\,$\mu$m sample}. Broadly speaking, the peak represents an average temperature for
the bulk of the dust. In the nearby Universe, more luminous galaxies are characterised by emission
from dust at higher temperatures which shifts the SED peak to shorter
wavelengths, explicitly shown in the SED peak--L$_{IR}$ relationship of
the \textit{local sample}; see also Sanders $\&$ Mirabel
(1996\nocite{1996ARA&A..34..749S}) and the peak-L$_{IR}$ variation in
the R08 SED templates, overplotted for comparison. 

The local and high redshift sources, especially objects that fall in
the LIRG and ULIRG luminosity class, occupy almost distinct
regions of this plot, signifying that for a given L$_{IR}$, high redshift
galaxies have on average colder SEDs that peak at longer wavelengths. 
Since the SED peak seems to evolve from the local to the high redshift
Universe, we might also expect it to evolve with redshift
within the \textit{70\,$\mu$m sample}, i.e. from z$\sim$0.1 to
$\sim$2. However, we are unable to examine this because, as mentioned earlier, the L$_{IR}$-z plane is not
sufficiently sampled. Under the circumstances, we rely on the more unbiased
comparison between local and high redshift sources by examining the
distribution in peak wavelength separately for the 3 luminosity classes
--- figure \ref{fig:peakdist}. SBs in both samples seem to be broadly
consistent, as the SB population in the \textit{70$\mu$m sample} is at quite a low mean redshift (z$\sim$0.2) and cosmic evolution in the
z$\sim$0 to z$\sim$0.2 interval is not expected to be
prominent. However, there are a few exceptions, with some
70\,$\mu$m-selected SBs being colder than local equivalents, but since these
sources are at low redshifts, this could be a selection effect due to the much
deeper MIPS survey. As described in section \ref{sec:limits}, this
argument does not apply to the higher redshift, higher luminosity
sources and hence any
differences in the types of objects picked up by the two surveys will be solely due to evolution.

Figures \ref{fig:LIRpeak} and \ref{fig:peakdist} show that, although
there is some overlap in the types of objects selected in the two surveys, it
seems that the SEDs of a substantial number of high redshift LIRGs and ULIRGs peak at
longer wavelengths than their local counterparts. Such a hypothesis was proposed in S08, where fitting locally-derived SED templates underestimated the
far-IR emission of a large fraction of high redshift objects. 
Also, it seems that the galaxies
in the \textit{70$\mu$m sample} span approximately the same range in SED peak,
whatever their luminosity class, in contrast to the narrower
local distribution, which notably shifts towards shorter wavelengths at higher luminosities.
Although the majority of local LIRGs peak at $\sim$75\,$\mu$m,
with a smaller fraction at $\sim$100\,$\mu$m (mean $\lambda_{peak}$$\sim$77\,$\mu$m), high redshift LIRGs span a
wide range in SED peak up to 160\,$\mu$m, with a mean of
$\sim$105\,$\mu$m. This effect is even more
pronounced in the ULIRGs, where the average peak in the SEDs shifts from
an average local value of $\sim$60\,$\mu$m to a high-redshift value of
$\sim$92\,$\mu$m.

\subsection{Submm colours}
\label{sec:submm}

In sections \ref{sec:70160} and \ref{sec:peak} we presented strong evidence
that the \textit{70\,$\mu$m sample} is not simply a redshifted version of sources in the local
Universe. What
naturally follows is a comparison with the SubMillimetre Galaxy (SMG)
population, comprised mostly of cold galaxies at high redshift. 
Due to the negative \textit{K}-correction and sensitivity limits, SMGs discovered in blank field surveys are mostly
in the 1.5$<$z$<$3.5 redshift range (e.g. Chapman et
al. 2005\nocite{2005ApJ...622..772C}), as the 350--850$\mu$m wavelength range corresponds
to 140--340$\mu$m $<$ $\lambda_{rest}$ $<$ 78--188$\mu$m, effectively
sampling near the SED peak for a wide range of sources. 

\begin{figure*}
\epsfig{file=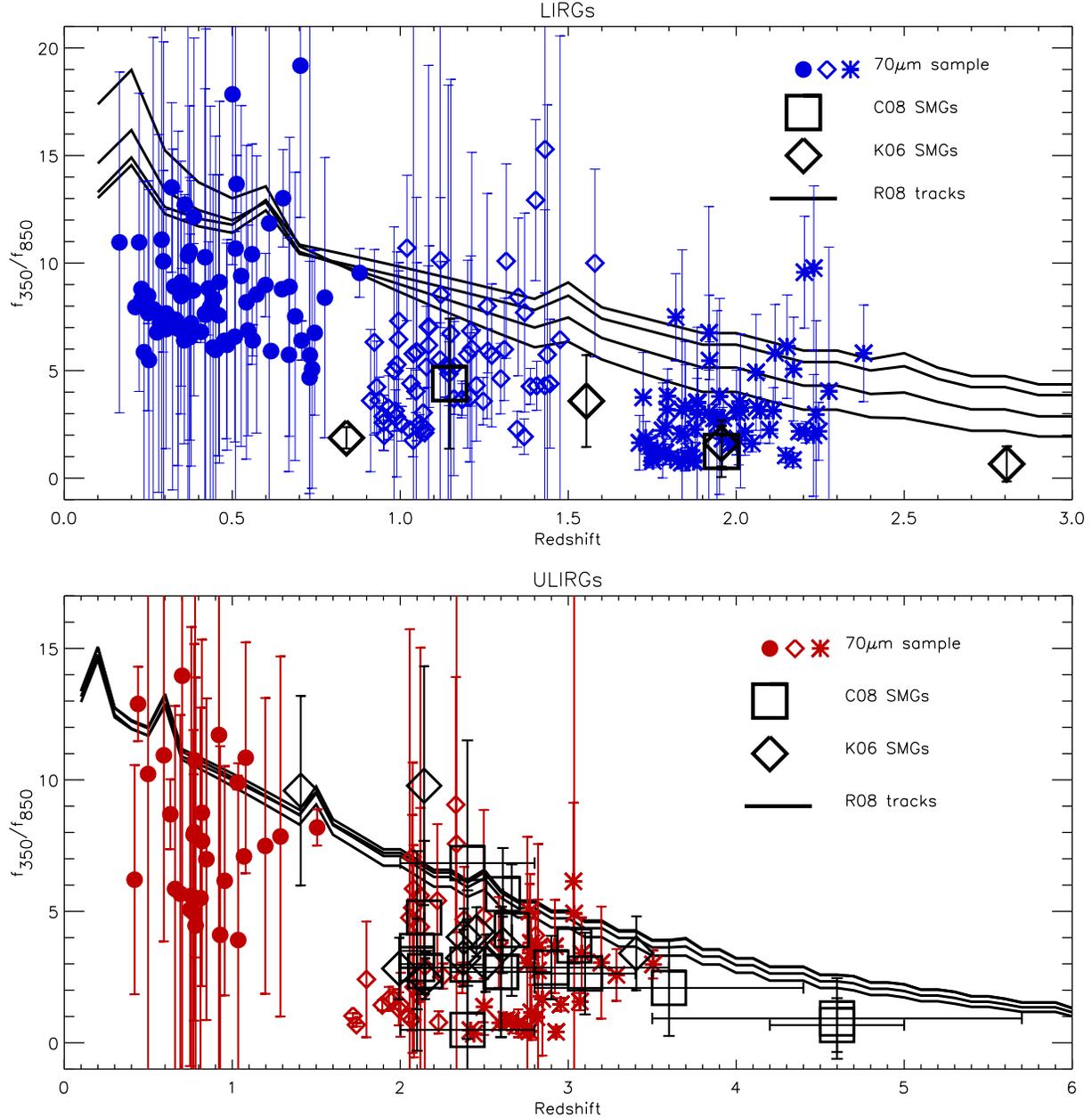,width=17cm}
\caption{Plot of the f$_{350}$/f$_{850}$ ratio versus redshift for
 LIRGs (top panel) and ULIRGs (lower panel). For reference the R08
 tracks are also included (solid black curves). The flux density
 ratios corresponding to the \textit{70\,$\mu$m sample} are
 predicted using the SK07 models. {\bf{Top panel}}:
 R08 LIRG tracks of log\,L$_{IR}$=11, 11.25, 11.5,
 11.75, all LIRG SMGs from Kovacs et al. (2006, K06) (large black
 diamonds) and Coppin et al. (2008, C08) (large black squares), all
 LIRGs in our sample at the original redshift z (blue filled
 circles), at a redshift z+0.7 (blue open diamonds) and at a redshift
 z+1.5 (blue asterisks) --- the change in redshift enables a comparison with the K06 and C08 LIRGs. 
{\bf{Lower panel}}: R08 ULIRG tracks of
 log\,L$_{IR}$=12, 12.25, 12.5, 12.75, all ULIRG SMGs from K06 (large
 black diamonds) and C08 (large black squares), the ULIRGs in our sample
 at the original redshift z (red filled
 circles), at a redshift z+1.3 (red open diamonds) and at a redshift
 z+2 (red asterisks) --- again the redshift change brings the
 70\,$\mu$m-detected ULIRGs to
 the redshift range of the K06 and C08 ULIRGs. The horizontal error bars denote errors on
 photometric redshifts in the C08 sample. The vertical error bars for the \textit{70\,$\mu$m
 sample} stem from the range of templates used to calculate
 f$_{350}$/f$_{850}$ (see section \ref{sec:methodology}). Note that for a few SMGs, the uncertainty in L$_{IR}$ could
 shift them to a higher luminosity class as some are very close to the
 upper luminosity limit. However, this does not alter the strong
 agreement in f$_{350}$/f$_{850}$ colour between SMGs and the 70\,$\mu$m
 sources, implying that we are detecting the lower redshift analogs of
 submm galaxies.}
\label{fig:submm350}
\end{figure*}

Using the SK07 fits, we predict 350 and 850$\mu$m flux
densities for the LIRGs and ULIRGs in the
\textit{70$\mu$m sample}, with uncertainties derived from the range of
best-matched templates (see section \ref{sec:methodology}) and plot the
evolution of the f$_{350}$/f$_{850}$ colour with redshift (figure
\ref{fig:submm350}). We compare with LIRG and ULIRG SMG data from Kovacs et al. (2006, hereafter K06) and
Coppin et al. 2008 (hereafter, C08). Also plotted are the R08 tracks, to represent the local IR galaxies. As the LIRGs in the \textit{70$\mu$m sample} are at an average redshift of
0.4, we predict the sub-mm flux densities they would
have if their redshifts were increased by 0.7 and 1.5, in order to bring them to
a similar redshift range as the K06 and C08 LIRGs (top panel). Similarly, we calculate what
submm flux densities we would expect for the ULIRGs (average z$\sim$0.8) if their redshift were increased by 1.3 and 2
(lower panel).

The majority of K06 and C08 LIRGs and ULIRGs fall below the corresponding R08 tracks, implying a lower f$_{350}$/f$_{850}$ ratio; for a given L$_{IR}$, sources with lower
f$_{350}$/f$_{850}$ are colder, as there is a shift in far-IR
emission, with more flux coming out at longer wavelengths. A large
number of sources from the \textit{70\,$\mu$m sample}
have similarly low f$_{350}$/f$_{850}$, again mostly below what is denoted by the R08
tracks, indicative of more cold dust emission, compared to what is seen locally.
The agreement in f$_{350}$/f$_{850}$ between
the SMGs and MIPS sources is a key result; it strongly suggests that the galaxies we detect at
70$\mu$m are the z$<$1 equivalents of the cold z$>$1 sources detected
in submm surveys.

\section{Discussion}
\label{sec:discussion}

\subsection{Cold Galaxies at z\,$\lesssim$\,1}

We find evidence for evolution in the dust/star-formation properties
with redshift, initially apparent through the f$_{160}$/f$_{70}$ colour and further
quantified by the rest frame wavelength of the SED peak. More
specifically, we show that a population of
LIRGs and ULIRGs at 0.1$<$z$<$2 appears to have a cold-dust-associated far-IR
excess not seen locally. 
Although we do identify many sources consistent with the kind of
objects detected in the local Universe, such as the well-studied M82 and
Arp220 types, our results establish a shift in the rest-frame
far-infrared SEDs of high redshift sources towards longer wavelengths. 
A simple description for this, is that the shapes of the far-infrared SEDs of many high redshift sources may resemble more closely those of
lower-luminosity galaxies locally. 

The strong cold component identified in many of our 70\,$\mu$m objects must relate
closely to conditions in the ISM, as it represents an increase in
far-IR/submm flux compared to local galaxies and hence must represent a change in dust
properties and/or star formation efficiency. For a given total infrared
luminosity, sources that are described by colder SEDs may consequently be
characterised by larger dust masses and/or higher
dust opacity, an increase in grain emissivity or a combination of
these. Another possibility is a change in the dust distribution, where a typical high-z LIRG/ULIRG may not be similar in structure
to the strongly nuclear-concentrated local examples; instead the dust
could be distributed over kpc scales,
consistent with measurements of their extent in CO (e.g. Tacconi et
al. 2006\nocite{}, Iono et al. 2009\nocite{}).

Hints for evolution have also emerged through other studies of high redshift, IR-selected
sources. For example, Rowan-Robinson et
al. (2004\nocite{2004MNRAS.351.1290R}, 2005\nocite{2005AJ....129.1183R})
report results from the European Large Area ISO Survey (ELAIS)
and the Spitzer Wide-area InfraRed Survey (SWIRE), which indicate that many IR-luminous
galaxies in the 0.15--0.5 redshift range are best fitted with cool
cirrus models. Examination of IR SEDs
for z$>$0.5 ULIRGs, has revealed a higher cool to hot dust ratio than
what is seen locally (Marcillac et al. 2006, Brand et al. 2008,
Chakrabarti $\&$ McKee 2008\nocite{2008ApJ...683..693C}). Sajina
et al. (2006\nocite{2006MNRAS.369..939S},
2008\nocite{2008ApJ...683..659S}) present results based on mid and
far-IR selected samples, which suggest that their sources are
better described by cooler temperatures and high far/mid-IR ratios in contrast to
local ULIRGs. Le Floch et al. (2005) find that ISOCAM-detected LIRGS at a median redshift of
z$\sim$0.7  are characterised by half solar metallicities which
could point to the IR SEDs being different from those of local
LIRGs. This behaviour is an interesting complement to the evidence for evolution from mid-infrared
spectroscopy and particularly the characteristics of Polycyclic Aromatic Hydrocarbon
(PAH) features. The aromatic bands in high redshift galaxies are found to
be similar in prominence and profile to the aromatic bands in local galaxies of lower
luminosity (e.g. Pope et al. 2008\nocite{}, Rigby et al. 2008\nocite{2008ApJ...675..262R}, Farrah et
al. 2008\nocite{2008ApJ...677..957F}), an important result as the PAH feature strengths in local galaxies are strongly
correlated with emission from cold dust (Haas et al. 2002\nocite{}, Bendo
et al. 2008\nocite{}). 

Predictions from several studies are also in agreement with the mounting evidence that the ISM, star-formation efficiency and
dust properties of many high redshift infrared galaxies significantly
diverge from those in the local universe. Chemical evolution models (Calzetti $\&$ Heckman 1999\nocite{1999ApJ...519...27C}, Pei, Fall $\&$
Hauser 1999\nocite{1999ApJ...522..604P}) predict a lower contribution in the UV SFR density and an increase in the average A$_V$ by a factor of 2--3 at z$\sim$1--2, an outcome also supported by observations in Efstathiou
$\&$ Rowan-Robinson (2003\nocite{2003MNRAS.343..322E}) and Le Floch et al. (2005\nocite{2005ApJ...632..169L}). Le Floch et
al. (2005) find a much stronger evolution in the infrared comoving
energy density than in the UV over the same redshift range, evidence that
dust reprocessing of optical light is more efficient at high redshift, but
also tying in with an increase in star formation efficiency (SFR
per unit cold gas mass) (Somerville, Primack and Faber 2001\nocite{2001MNRAS.320..504S}). 

The observations and results outlined above are likely to have considerable impact in accurately modelling the far-IR
and submm background, as well as pinpointing the evolution in the infrared luminosity function, where the cold SED component may play a more significant role than previously
thought. We expect that cold galaxies such as those we identify here, will be discovered in
abundance by the \textit{Herschel Space Observatory} (Pilbratt et al. 2004\nocite{2004AAS...204.8101P}). The capabilities of Herschel's Spectral and Photometric
Imaging Receiver (SPIRE, Griffin et al. 2007\nocite{2007AdSpR..40..612G}) and Photodetector
Array Camera (PACS, Poglitsch et al. 2008\nocite{2008SPIE.7010E...3P}) will provide detailed measurements in the
60--700\,$\mu$m region. This will allow accurate mapping of
previously unexplored territories such as the nature of the main
contributors to the cosmic infrared background and ISM properties at
high redshift (Harwit 2004\nocite{2004AdSpR..34..568H}).  
The question as to whether a large fraction of IR-luminous galaxies at
high redshift are cold, will explicitely be answered since the SED peak will be
entirely defined at least up to redshift z$\sim$3, enabling precise
temperature measurements. In addition, with
\textit{Herchel}'s capabilities the nature and properties of dust at high
redshift and its evolution through cosmic time will be accurately
traced.

\subsection{The submm/IR connection}
\label{sec:submmir}
SMGs, which are predominantly in the ULIRG and HyLIRG regime, are
well-established as the most dust-rich galaxies in the Universe. Comparisons between the SEDs of SMGs and those of local and lower redshift
(z\,$<$\,1.5) populations are therefore essential in the context of galaxy
evolution, particularly evident when modelling the behaviour of the IR and submm source counts, which necessarily
requires the assumption of a SED shape. 
If SMGs are the high redshift analogs of local
warm IR galaxies, the high
submm counts on the faint end (with the limits of present surveys) could only be reproduced by evoking strong evolution well
beyond z$\sim$3-4, as a warm SED moves the peak to shorter wavelengths, reducing the submm
flux for a given L$_{IR}$. This goes against the predictions of hierchical
galaxy formation models which anticipate a
downturn in space density of IR-luminous sources with Star
Formation Rates (SFRs) $>$100\,M$_{\odot}$/yr by z$\sim$3-4. Instead, if
submm emission arises from a significantly cooler SED, submm counts
could be fit with galaxies of more
moderate SFRs (Kaviani, Haehnelt and Kauffman
2003\nocite{2003MNRAS.340..739K}). Predictions by Efstathiou et al. (2000\nocite{2000MNRAS.319.1169E}), Rowan-Robinson (2001\nocite{2001ApJ...549..745R}) and Efstathiou $\&$
Rowan-Robinson (2003\nocite{2003MNRAS.343..322E}) also support this argument, suggesting that submm emission in high
redshift galaxies could predominantly arise from cold dust and
proposing that the quiescent cirrus component undergoes the same strong luminosity evolution as the
starburst component. 

Up to now, studies of SMG SEDs converge on the idea that they are
predominantly fit by cooler cirrus SED templates (e.g. by Chapman et al. 2005\nocite{2005ApJ...622..772C}, Kovacs et al. 2006\nocite{2006ApJ...650..592K}, Pope et
al. 2006\nocite{2006MNRAS.370.1185P}, Coppin et al. 2008\nocite{2008MNRAS.384.1597C}, Clements et
al. 2008\nocite{2008MNRAS.387..247C}). 
Submm studies of high redshift ULIRGs and HyLIRGs, have shown that their
intrinsic dust temperatures are distributed around a mean of
T$\sim$35\,K (Chapman et al. 2005\nocite{}, Kovacs et
 al. 2006\nocite{}, Pope et al. 2006\nocite{}, Coppin et
 al. 2008\nocite{}). Huynh et al. (2007\nocite{2007ApJ...659..305H}) report
very low typical dust temperatures (T=21--33\,K) for a sample of L$>$5$\times$10$^{11}$
 low-z SMGs. Interestingly, this corresponds to the
 range in temperatures seen in the local SBs and LIRGs described in
 D00 and it is in contast to the higher temperatures observed 
in local ULIRGs --- the Farrah et al. (2003\nocite{2003MNRAS.343..585F}) local ULIRG sample is fit by
 T\,$\sim$\,42\,K; see also Klaas et al. (1997\nocite{1997A&A...325L..21K}) and Soifer et
 al. (1984)\nocite{1984ApJ...283L...1S} where the SEDs of Arp220 and
 NGC6240 are fit by modified blackbodies of temperature 47\,K and 42\,K
 respectively. 
Note that even if submm studies preferentially detect cold sources (e.g. Blain et
al. 2004\nocite{2004ApJ...611...52B}), many of these have
total infrared luminosities comparable to the most energetic local
galaxies, supporting the above evidence that a large number of
high-redshift IR-luminous objects are colder than their local
analogs. In fact the same conclusions are reached by Sajina et
al. (2008) for a sample of high-redshift
mid-IR selected sources, showing
that this is not simply an artifact of the submm selection. 

Our results have established a connection between the far-IR and submm, showing that high redshift SMGs likely share similar properties
with the lower redshift galaxies selected in \textit{Spitzer}/MIPS
surveys. We find that the predicted submm colours for the
70\,$\mu$m-detected LIRGs and ULIRGs when redshifted to the
1.5$<$z$<$3.5 range, trace the same parameter space as
the actual colours of submm-detected sources in the same luminosity and
redshift range. This suggests that we have identified the missing link between SCUBA and \textit{Spitzer}/MIPS
galaxies, i.e. a population at intermediate redshift (z\,$\sim$\,1), detected by
\textit{Spitzer} but with the properties commonly seen in SCUBA galaxies.

\section{Conclusions}

We have examined the properties of a large sample of 0.1$<$z$<$2
IR-luminous sources selected at
70\,$\mu$m, corresponding to rest-frame 23--63\,$\mu$m, a
selection which should strongly favour starforming galaxies. 
Using available photometry in the 8--160$\mu$m range, we fit the
 Siebenmorgen $\&$ Krugel (2007, SK07) dust templates in order
 to calculate the total infrared luminosity and rest-frame wavelength of
 the SED peak, finding the former to be mainly in the 10$^{10}$--10$^{14}$
 L$_{\odot}$ range and the latter at an average of $\sim$105\,$\mu$m for the LIRGs
 and $\sim$92\,$\mu$m for the ULIRGs. As the main
 aim of this study is to compare our results with local sources of
 equivalent luminosity, we use data from the local \textit{IRAS} Bright Galaxy
 Sample, again fitting the photometry with the SK07 models. The
 locally-derived SED templates of Chary $\&$ Elbaz (2001, CE01)
 and Rieke et al. (2008, R08) are also used as an additional comparative
 tool. We pay particular attention to any sources of bias that
 could affect our results and find nothing that could potentially
 predispose the outcome of our analysis.

Our major findings are:
\begin{itemize}
\item By examining the f$_{160}$/f$_{70}$ colour of the \textit{70$\mu$m sample}, we find that a large fraction of LIRGs and ULIRGs
      have higher f$_{160}$/f$_{70}$ than what is
      defined by the empirical CE01 and R08 local galaxy SEDs, by at least a factor of
      2 and up to 10 in extreme cases. As the f$_{160}$/f$_{70}$ continuum slope is
      directly related to the position of the SED peak wavelength, non-trivial differences between local
      and high redshift IR populations are implied.
\item For a given total infrared luminosity,
      a large fraction of the LIRG and ULIRG SEDs in the \textit{70\,$\mu$m
      sample} peak at wavelengths longer than 90\,$\mu$m, indicative
      of a significant cold-dust-associated far-IR excess not detected in local sources of
      comparable luminosity. This behaviour is directly
      linked to evolution in dust and/or star-formation
      properties from the local to the high redshift universe. 
\item We predict the f$_{350}$/f$_{850}$ colours that the
      70$\mu$m-selected LIRGs and ULIRGs would have if found at the average redshift of
      SMG populations (1\,$<$\,z\,$<$\,3) and find that they are on
      average equivalent to the observed SMG colours. This result suggests that we have
 potentially identified the z$<$1 equivalents of the cold z$>$1 sources discovered in blank
field submm surveys. 
\end{itemize}

The strong evidence that we present here, drives the answer to the question that we
defined at the start, i.e. whether there is a substantial number of
infrared galaxies at high redshift with properties divergent from those
of their local counterparts, to be yes. More specifically, our results
have shown that these sources being colder rather than warmer. The
emergent conclusion is that the evolution of the IR-luminous population with redshift
is towards a larger dust mass or higher dust opacity and emissivity, a more
extended dust distribution, higher star-formation efficiency or a
combination of all. The imminent \textit{Herschel} observational
programmes will be key in identifying and disentangling the main processes responsible.

\section*{Acknowledgments}
This work is based on observations made with the Spitzer Space
Telescope, operated by the Jet Propulsion Laboratory, California
Institute of Technology, under NASA contract 1407 and partially
supported by JPL/Caltech contract 1255094 to the University of
Arizona. KC is grateful for STFC fellowship support. Special thanks to David Alexander for insightful discussions.

\bibliographystyle{mn2e}
\bibliography{references}

\label{lastpage}

\end{document}